\documentclass[11pt]{article}
\usepackage{fullpage}

\usepackage[utf8]{inputenc}

\usepackage[normalem]{ulem}
\usepackage[fleqn]{mathtools}
\usepackage{bm}
\usepackage[inline]{enumitem}
\usepackage{amsfonts,amssymb,amsthm,hyperref,algorithm,eqparbox,array,ragged2e,stmaryrd,bbm} 
\usepackage{physics}
\usepackage{xspace}
\usepackage{tikz}
\usepackage{xcolor}
\usepackage{tabularx}
\usepackage{ifthen}
\usepackage{pgfmath-xfp}
\usetikzlibrary{positioning,decorations.pathreplacing,calligraphy, arrows.meta}
\pgfmathsetseed{3434}
\usepackage{algorithm}
\usepackage{algpseudocode}

\usepackage{thm-restate}
\usepackage{enumitem}
\usepackage[capitalise,nameinlink,noabbrev]{cleveref}

\definecolor{MidnightBlue}{rgb}{0.1, 0.1, 0.44}
\hypersetup{
	colorlinks=true,
	urlcolor=MidnightBlue,
	linkcolor=MidnightBlue,
	citecolor=MidnightBlue,
	unicode
}

\newtheorem{theorem}{Theorem}
\newtheorem{lemma}[theorem]{Lemma}

\newtheorem{claim}[theorem]{Claim}

\newtheorem{fact}[theorem]{Fact}

\theoremstyle{definition}
\newtheorem{definition}[theorem]{Definition}

\theoremstyle{remark}

\makeatletter
\def\th@example{%
  \thm@notefont{}
  \normalfont 
}
\def\th@definition{%
  \thm@notefont{}
  \normalfont 
}
\makeatother

\theoremstyle{example}

\newcommand{\deftext}[1]{\emph{#1}}

\renewcommand{\setminus}{\smallsetminus}

\newcommand{\Cl}{\mathsf{Cl}}

\renewcommand{\H}{\operatorname{H}}

\newcommand{\Hinf}{\textbf{H}_{\infty}}
\newcommand{\Dinf}{\textbf{D}_{\infty}}

\DeclareMathOperator{\Exp}{\mathbb{E}}

\renewcommand{\epsilon}{\varepsilon}

\newcommand{\fix}{\textsf{\upshape fix}}

\newcommand{\poly}{\mathrm{poly}}

\newcommand{\codim}{\mathrm{codim}}

\newcommand{\Search}{\mathrm{Search}}

\newcommand{\newmeasure}[2]{\newcommand{#1}{{\textup{\sffamily #2}}\xspace}}
\newmeasure{\C}{C}

\newcommand{\ZO}{\{0,1\}}

\let\oldsum=\sum 

\RenewDocumentCommand{\sum}{e{_^}}{ 
  \vphantom{\oldsum_{.}} 
  \mathop{\smash{\oldsum 
    \IfValueT{#1}{_{#1}} 
    \IfValueT{#2}{^{#2}}
  }}
}
\let\oldprod=\prod 

\RenewDocumentCommand{\prod}{e{_^}}{ 
  \vphantom{\oldprod_{.}} 
  \mathop{\smash{\oldprod 
    \IfValueT{#1}{_{#1}} 
    \IfValueT{#2}{^{#2}}
  }}
}

\ifthenelse{1>0}{
\newcommand{\artur}[1]{\textcolor{orange}{[Artur: #1]}}
\newcommand{\mika}[1]{\textcolor{blue}{[Mika: #1]}}
\newcommand{\weiqiang}[1]{\textcolor{red}{[Weiqiang: #1]}}
\newcommand{\dima}[1]{\textcolor{green!33!black}{[Dima: #1]}}

}{
\newcommand{\artur}[1]{}
\newcommand{\mika}[1]{}
\newcommand{\weiqiang}[1]{}
\newcommand{\dima}[1]{}
\newcommand{\ziyi}[1]{}
\newcommand{\gilbert}[1]{}
\newcommand{\yaroslav}[1]{}
}

\renewcommand{\phi}{\varphi}

\newcommand{\Err}{\textsc{Error}}

\title{
    \huge
    Searching for Falsified Clause in Random $(\log{n})$-CNFs is Hard for Randomized Communication
}

\author{%
    \begin{tabularx}{4cm}{c}
      Artur Riazanov \\ \textit{EPFL} \\ {\small artur.riazanov@epfl.ch} \\ \\
      Dmitry Sokolov \\ \textit{EPFL} \\ {\small sokolov.dmt@gmail.com}
    \end{tabularx}
    \and
    \hspace{1cm}
    \and
    \begin{tabularx}{4cm}{c}
      Anastasia Sofronova \\ \textit{EPFL} \\ {\small anastasiia.sofronova@epfl.ch} \\ \\
      Weiqiang Yuan \\ \textit{EPFL} \\ {\small weiqiang.yuan@epfl.ch}\\
    \end{tabularx}
}

\begin{document}

\maketitle

\begin{abstract}
    We show that for a randomly sampled unsatisfiable $O(\log n)$-CNF over $n$ variables the randomized
    two-party communication cost of finding a clause falsified by the given variable assignment is
    linear in $n$.    
\end{abstract}

\section{Introduction}

This paper studies the communication complexity of
\emph{Falsified Clause Search Problem}.

\begin{definition}[\cite{LNNW95}]
    Let $X, Y$ be two disjoint sets of boolean variables and $\varphi$ be a CNF formula over the
    variables $X \sqcup Y$. We define \deftext{Falsified Clause Search Problem} or $\Search_{\varphi}$
    associated with formula $\varphi$ in the following way:
    \begin{itemize}[itemindent = 1cm, itemsep = 0cm, topsep = 0.1cm]
        \item[\textit{input:}] a pair $(x, y) \in \{0, 1\}^{X} \times \{0, 1\}^{Y}$;
        \item[\textit{output:}] a clause $C \in \varphi$ that is violated by the input $(x, y)$.
    \end{itemize}
\end{definition}

Communication lower bounds for search problems have applications in many areas of complexity theory. We
consider two areas that are the most relevant and explain the applicability of communication lower
bounds.

\paragraph{Proof complexity.} This area of complexity theory studies how hard it is to prove that a given
formula $\varphi$ is unsatisfiable; in other words, what is the length of the shortest proof in a certain
proof system. Lower bounds for the proof systems often correspond to lower bounds on a run-time of
SAT-solvers, and there are intricate connections to other areas of complexity theory, such as, for
example, circuit complexity.

There is a general framework for obtaining lower bounds on the length of the shortest proofs via
communication. Suppose that, for an unsatisfiable CNF formula $\varphi$, we divide the variables into two
disjoint groups $X$ and $Y$ in an arbitrary way. For a fixed proof system $\mathfrak{C}$ we can try to
transform efficient proof of $\varphi$ into an efficient communication protocol for
$\Search_{\varphi}$. A lower bound on the communication complexity of
$\Search_{\varphi}$ then implies a lower bound on the length of a proof of $\phi$ in $\mathfrak{C}$.

This framework seems to originate from \cite{LNNW95}. Following this reduction, lower bounds for many
different proof systems were obtained, for example: tree-like Cutting Planes
\cite{IPU94, HN12, RNV16, BW24}, tree-like Threshold proof system \cite{BPS07}, tree-like
$\mathrm{Res}(\oplus)$ \cite{IS20}, etc. \cite{HN12, GP18-cbs}. Depending on the communication model,
even dag-like proofs can be analyzed via this framework
\cite{Krajicek97, Pudlak97, HrubesP17, FPPR22, GGKS20, S24}.

The lower bounds that can be achieved via this technique depend on the power of the communication model:
the more powerful model we consider, the bigger class of proof systems we get the lower bound for. The
choice of the formula $\varphi$ is important here as well, in a sense that we need to be able to show the
lower bound on the communication complexity of $\Search_{\varphi}$. Typically, $\varphi$ is artificially
built for this purpose. In this paper, we show a communication lower bound for the natural class of
formulas (without usage of ad hoc constructions) that is a candidate for being hard for all propositional
proof systems.

\paragraph{Circuit complexity.} Natural embedding of $\Search_{\varphi}$ into a monotone
Karchmer--Wigderson relation \cite{KW90, Razborov90} gives us the opportunity to use it for proving lower
bounds for the monotone models of computation. From communication lower bounds, strong results are known
for monotone formulas \cite{RPRC16}, monotone circuits \cite{GGKS20, LMMPZ22}, monotone span programs
\cite{RPRC16, PR18-null}, etc. Communication is also the main instrument for showing separation between
those models \cite{PR18-null, GKRS19}, and trade-off results \cite{dRFJNP24, MGKS25}. These type of
results are based on ad hoc constructions of the formulas $\varphi$. Namely, $\varphi$ is designed in
order to able to show communication lower bound.

\subsection{Random CNF}

To be more precise we start with the definition of random CNF formulas.

\begin{definition}
    \label{def:random-formulas}%
    Let $\mathfrak{F}(m, n, \Delta)$ denote the distribution of random $\Delta$-CNF on $n$ variables
    obtained by sampling $m$ clauses (out of the $\binom{n}{\Delta} 2^{\Delta}$ possible clauses)
    uniformly at random with repetitions.
\end{definition}

The famous result of Chv{\'{a}}tal--Szemer{\'{e}}di says that if we pick a formula from this distribution
with the proper parameters, the resulting formula will be unsatisfiable with high probability. 

\begin{theorem}[Chv{\'{a}}tal--Szemer{\'{e}}di, \cite{ChvSem88}]
    \label{th:unsat-ch-sz}%
    For any $\Delta \ge 3$ whp $\bm{\varphi} \sim \mathfrak{F}(m, n, \Delta)$ is unsatisfiable if $m \ge
    \ln 2 \cdot 2^{\Delta} n$.
\end{theorem}

These types of distributions appear not only in most of the areas in computer science, but in
general mathematics and physics as well \cite{MezParZec02}. An interesting application is due to Feige
\cite{Feige02}, who conjectured the following statement: no polynomial time algorithm may \emph{prove}
whp the unsatisfiability of a random $O(1)$-CNF formula with arbitrary large constant clause
density. Assuming Feige's conjecture, it is known that some problems are hard to approximate: vertex
covering, PAC learning DNFs~\cite{HardnessLearningDNFs}, etc.

As a candidate to be hard to refute in all proof systems, random CNFs are actively studied and lower
bounds are known for many different proof systems
\cite{Grig01, BKPS02, AR03, Alekhnovich11, SofronovaS22}. Recent developments in this direction utilize
the connection between proof complexity of $\varphi$ and communication complexity of
$\Search_{\varphi}$. In particular, lower bounds for the Cutting Planes proofs of random $O(\log n)$-CNF
\cite{HrubesP17, FPPR22, S24} follow this strategy. However, these results only consider lower bounds on
deterministic \emph{dag-like communication complexity} of $\Search_{\varphi}$ based on random $O(\log
n)$-CNF.

In this paper, we analyse the randomized tree-like communication of this problem that is incomparable
with deterministic dag-like communication. This is a natural problem in a natural model, which also
provides a way to explore how techniques used for structured formulas might extend to more typical
instances like random CNFs. The main result is the following.

\begin{theorem}
    \label{th:main-random}%
    Let $c > 0$ be a large enough constant, $n > 0, \Delta \ge c \log n, m = O(n 2^{\Delta})$. If
    $\bm{\varphi} \sim \mathfrak{F}(m, n, \Delta)$ and $\bm{X}, \bm{Y} \subseteq [n]$ is a partition of
    variables that is taken uniformly at random, then whp over choice of $\bm{\varphi}$ and partition
    $\bm{X}, \bm{Y}$ the randomized communication complexity of $\Search_{\bm{\varphi}}$ is $\Omega(n)$.
\end{theorem}

\subsection{Prior Results and Technique}

For several types of formulas $\varphi$, the randomized communication complexity of $\Search_{\varphi}$
is well-studied. The approach for proving such bounds is the reduction of \emph{Unique Disjointness}
function to $\Search_{\varphi}$. The main success in this direction is the reduction based on
\emph{critical block sensitivity} \cite{HN12, GP18}, we also include some earlier results, though there
is some difference in the technique \cite{BPS07}. More precisely, for this technique one should assume
that $\varphi = \psi \circ g$ (we take some formula $\psi$ and, in place of each variable, we substitute
a carefully chosen gadget $g$ with fresh variables). Assuming that $\Search_{\psi}$ has critical block
sensitivity $m$ (that is a generalization of the block sensitivity measure), it is possible to reduce
instances of unique disjointness of size $\poly(m)$ to $\Search_{\varphi}$.

The general framework for working with such formulas of $\psi \circ g$ is called \emph{lifting},
and the idea is to ``lift'' the hardness of $\psi$ with respect to another complexity measure to
communication complexity via gadget. Lifting can be based on the other complexity measures as well. For
example, it can also be implemented for randomized decision tree complexity instead of critical block
sensitivity \cite{GJPW18}; however, this method requires the lower bound on the randomized decision tree
complexity, which might be non-trivial, especially in case of $\Search_{\varphi}$ problem. Such lower
bound is known for Tseitin formulas \cite{GJW18}, together with \cite{GJPW18} it yields the lower bound
for randomized communication complexity of $\Search$ for Tseitin formulas lifted by Inner Product.

The notable exception here is the lower bound on $\Search$ problem for \emph{Binary Pigeonhole Principle}
(BPHP) \cite{ItsyksonR21}. These formulas are not lifted, however the proof is also the reduction
of Unique Disjointness to the $\Search$ problem. This reduction based on the inner symmetry of BPHP.

A different kind of proof of a $\Search_{\varphi}$ lower bound was given by Yang and Zhang~\cite{YZ24}
(based on \cite{WYZ23, YZ24}), who prove a lower bound for a weak version of BPHP. In contrast to the
previous works this one is not a reduction from Unique Disjointness. Instead, they directly apply the
structure versus randomness framework from the lifting literature \cite{GLMWZ16, GJPW18} to the potential
protocol that computes $\Search_{\varphi}$.

Our proof of \cref{th:main-random} combines the approach of \cite{GLMWZ16, GJPW18, YZ24} with the
analysis of expander graphs via closure argument that was developed for proof complexity purposes in
\cite{AR03, ABRW04}. However, we use the iterative construction of the closure from \cite{S20}. In part,
this is also inspired by \cite{Condensation}.  

More precisely, the proof of our result is based on the following steps.
\begin{enumerate}
    \item Following \cite{HrubesP17, FPPR22, S24} we divide variables between Alice and Bob uniformly at
        random. 
    \item Following the line of work on lifting of randomized decision trees \cite{GLMWZ16, GJPW18, YZ24,
        QuantumCommunicationTFNP} we show that every communication protocol can be converted into a more
        structured one, a so-called subcube-like protocol. In such a communication protocol, each
        rectangle is a product of two sets with some bits fixed and the remaining pseudorandom.
    \item Due to the nature of our random CNFs, the invariant that all clauses contain pseudorandom variables is not strong enough on its own. 
        $\Search$ problem still might become trivial early on in communication protocol; for example, if the contradiction could be narrowed down to a small set of clauses. To avoid this
        problem, we use the closure trick \cite{AR03, ABRW04, S20, Condensation}, that allows us to maintain
        expansion property on the pseudorandom part of the graph.
    \item Following \cite{QuantumCommunicationTFNP}, we show that the number of fixed bits in each
        rectangle is at most $O(d / \varepsilon)$ if we allow error $\varepsilon$, where $d$ is the
        communication complexity of the original protocol. 
\end{enumerate}

In addition, we show the better error bound dependency on the protocol depth $d$ than
in~\cite{QuantumCommunicationTFNP}. We give a more refined analysis of the conversion to the subcube-like protocols. More precisely, we
show that the number of fixed bits in each rectangle is $O(d)$ even when we allow for the $\exp(-d)$
error.

\section{Notation and Tools}
We denote the standard binary entropy function by $\H(p) \coloneqq p \log (1 / p) + (1 - p) \log (1 / (1 - p))$.

\begin{definition}
    A bipartite graph $G = (L, R, E)$ is called an $(r, \Delta, \alpha\Delta)$-expander, if all vertices
    in $L$ have degree at most $\Delta$ and for any set $S \subseteq L$ such that $|S| \leq r$ it holds
    that $|N(S)| \geq \alpha \Delta |S|$, where $N_G(S)$ denotes the set of neighbours of $S$ in $G$ (we
    omit the subscript if the graph is clear from the context).
\end{definition}

With a CNF formula $\varphi$ over $n$ variables and with $m$ clauses we associate a graph $G_{\varphi} \coloneqq
([m], [n], E)$ in a natural way: $(i, j) \in E$ iff the $i$-th clause contains the $j$-th variable. The following
Lemma gives us some useful properties of underlying graphs of random CNFs. It follows from a standard
computation, which was featured, for example, in \cite[Lemma~A.2]{S24}.

\begin{lemma}
\label{lm:exp-parameters}
    Let $n > 0$, $\eta > 0$ be an arbitrary constant, $\Delta = c \log{n}$, for a large enough constant
    $c$ depends on $\eta$, $m = O(n2^{\Delta})$. Let $G \coloneqq ([m], [n], E)$ be a bipartite graph,
    such that each $i \in [m]$ choses $\Delta$ neighbours uniformly at random over $\binom{n}{\Delta}$
    possibilities. Then $G$ is an $(r, \Delta, (1 - \eta) \Delta)$-expander for $r =
    \Omega\left( n / \Delta \right)$.
\end{lemma}

Instead of working directly with randomised communication, we use the equivalent characterisation through
distributional communication complexity. That is, we prove a lower bound against deterministic protocols
that achieve error $\varepsilon$ with respect to a certain distribution on inputs (here we use uniform
distribution), and a lower bound against randomised protocols that achieve error $\varepsilon$
follows. Below, ``communication protocol'' refers to a deterministic communication protocol.

\section{Refuting Bipartite CNFs}
In this section we mainly prove a special ``bipartite'' case of \cref{th:main-random}. We show in
\cref{sec:random_from_bipartite} that it actually implies the general case.

\begin{restatable}{theorem}{TheoremMainBipartite}
    \label{thm:bipartite-main}
    Let $\alpha > 0$ be an absolute constant. Let $G_1 \coloneqq ([m], [n], E_1), G_2 \coloneqq ([m],
    [n], E_2)$ be two $(r, \Delta, \alpha \Delta)$-expanders, and $X, Y \coloneqq \{0, 1\}^n$. For each
    $i \in [m]$, let $C_i$ be a disjunction of variables in $\{ x_j\mid j \in N_{G_1}(i)\} \cup \{
    y_j\mid j\in N_{G_2}(i)\}$ with arbitrary signs. Then for every communication protocol $\Pi\colon
    \ZO^{n}\times \ZO^n\to [m]$ of depth $d$ at most $O(\Delta r)$:
    \[
        \Pr_{\substack{(\bm{x}, \bm{y}) \sim X \times Y\\\bm{i}\sim\Pi(\bm{x}, \bm{y})}}
        [C_{\bm{i}}(\bm{x}, \bm{y}) = 0] \le d \cdot 2^{-\Omega(\Delta)} + \exp(-d).
    \]
\end{restatable}

This section is organized as follows. In~\Cref{sec:random_from_bipartite}, we derive
\Cref{th:main-random} from \Cref{thm:bipartite-main}. In~\Cref{sec:machinery}, we formally define
subcube-like protocols and provide necessary tools. In~\Cref{sec:subcube_from_general}, we give a more
refined analysis of the conversion from general protocols to subcube-like ones
in~\cite{QuantumCommunicationTFNP}. In~\Cref{sec:lower_bound_subcube}, we show the hardness of
$\Search_{\varphi}$ against subcube-like protocols when the underlying graphs are good
expanders. Finally, in~\Cref{sec:proof_bipartite}, we put everything together and
derive~\Cref{thm:bipartite-main}.

\subsection{Deriving \cref{th:main-random} from \cref{thm:bipartite-main}}
\label{sec:random_from_bipartite}
The main part of the argument that reduces the general case to the bipartite is a clean-up lemma
essentially saying that incurring a small error we can treat the general case as bipartite. Similar
arguments have been made in \cite{HrubesP17, FPPR22, S24}.

Let $\varphi = \bigwedge_{i \in [n]} C_i$ be a $\Delta$-CNF with the set of variables $[n]$. Let $A
\sqcup B = [n]$ be a partition of the variables. Let $G_A \coloneqq ([m], A, E_A)$ and $G_B \coloneqq
([m], B, E_B)$ be the graphs with edges connecting a clause with all variables from one of the sets
mentioned in it. Let $\Err_A \subseteq [m]$ and $\Err_B \subseteq [m]$ be the sets of clauses with degree
exceeding $(1-\delta)\Delta$ in $G_A$ and $G_B$ respectively. It means that clauses from $[m] \setminus
(\Err_A \cup \Err_B)$ have at least $\delta \Delta$ variables from $A$ and $B$. We then say that $(A,B)$
is \emph{$\delta$-good partition for $\varphi$} if
\begin{enumerate}
    \item $\Pr\limits_{\bm{x} \sim \{0, 1\}^A}[\forall i \in \Err_A\text{ we have } C_i(\bm{x}, \cdot)
        \equiv 1] \ge 1 - 2^{-\Omega(\Delta)}$. \label{item:whp-satA}
    \item $\Pr\limits_{\bm{y} \sim \{0, 1\}^B}[\forall i \in \Err_B\text{ we have } C_i(\cdot, \bm{y})
        \equiv 1] \ge 1 - 2^{-\Omega(\Delta)}$. \label{item:whp-satB} 
    \item $G_A - \Err_A - \Err_B$ and $G_B - \Err_A - \Err_B$ are $(r, \Delta, \delta \Delta /
        2)$-expanders, where $r = \Omega(n/\Delta)$. \label{item:whp-expander}
\end{enumerate}
In this definition we assume that $\delta$ is an absolute constant and hidden constants depend on it.

\begin{lemma}
\label{lem:partition}
    Let $\bm{\varphi} \sim \mathfrak{F}(m, n, \Delta)$ with $\Delta = c \log n$ and $m = \alpha 2^\Delta n$,
    where $c, \alpha > 0$ are constants, and $c \ge 40$. Let $\bm{X},\bm{Y}$ be a uniformly random
    partition of $[n]$. Then whp $(\bm{X},\bm{Y})$ is a $\delta$-good partition for $\bm{\varphi}$ for any
    $\delta \le 1 / 10$.
\end{lemma}
We defer the proof of this lemma to \cref{sec:proof-of-partititon}.

\begin{proof}[Proof of \cref{th:main-random}]
    Applying \cref{lem:partition}, we get that the variable partition $\bm{X} \sqcup \bm{Y} = [n]$ is
    $1 / 10$-good wrt $\bm{\varphi}$. Let $G_1 \coloneqq G_{\bm{X}} - \Err_{\bm{X}} - \Err_{\bm{Y}}$,
    $G_2 \coloneqq G_{\bm{Y}} - \Err_{\bm{X}} - \Err_{\bm{Y}}$. Note that the left parts of these graphs
    have equal size. We can add dummy variables to the right parts of these graphs to make them equal for
    the simplicity of notation.

    By \cref{lem:partition}, the probability over $(\bm{x}, \bm{y}) \sim \{0, 1\}^{\bm{X}} \times \{0,
    1\}^{\bm{Y}}$ for the $\Err_{\bm{X}}$ or $\Err_{\bm{Y}}$ to not be immediately satisfied is
    $2^{-\Omega(\Delta)}$. This means that if we consider a protocol for the problem
    $\Search_{\bm{\varphi}}$ for the variable partition $\bm{X} \sqcup \bm{Y}$ with the probability of
    success $\varepsilon$, we can reinterpret it as a protocol for $G_1$ and $G_2$ with the probability
    of success at least $\varepsilon - 2^{-\Omega(\Delta)}$.

    We apply \cref{thm:bipartite-main} with $\alpha \coloneqq \frac{1}{20}$. Since $r \Delta$ can be as
    large as $\Omega(n)$ by \cref{lm:exp-parameters}, we can pick the constants depending on $\alpha$
    such that the probability from \cref{thm:bipartite-main} is less than $\frac{1}{100}$. Then the
    probability of success for the problem from \cref{th:main-random} is less than $\frac{1}{100} +
    2^{-\Omega(\Delta)}$, and the theorem follows.
\end{proof}

\subsection{Density Restoring Machinery}\label{sec:machinery}


Every communication protocol $\Pi$ can be seen as a tree (not necessarily binary).
Let $\mathcal{N}(\Pi)$ denote the set of all nodes in $\Pi$.
Each node $v\in \mathcal{N}(\Pi)$ is associated with a rectangle, denoted $R_v=X_v\times Y_v$.

\begin{definition}[Min-entropy]
    For a random variable $\bm{x}$, let $\Hinf(\bm{x}) = \min_x \log{\frac{1}{\Pr[\bm{x} = x]}}$.
\end{definition}

\begin{definition}[Spread variables]
    Let $\bm{x} \in \ZO^n$ be a random boolean vector. We say $\bm{x}$ is $\gamma$-\deftext{spread} if
    for every $I \subseteq [n]$ we have $\Hinf(\bm{x}_I)\ge \gamma|I|$. 
\end{definition}

\begin{definition}[Structured variables]
    Let $\bm{x} \in \ZO^n$ be a random boolean vector and $I\subseteq [n]$. We say $\bm{x}$ is $(I,
    \gamma)$-\deftext{structured} if there exists some $a_I\in \ZO^I$ such that
    \begin{itemize}
        \item $\Pr[\bm{x}_I = a_I] = 1$;
        \item $\bm{x}_{[m]\setminus I}$ is $\gamma$-spread.
    \end{itemize}
\end{definition}

\begin{definition}[Subcube-like rectangle]
    A rectangle $R = X \times Y \subseteq \{0, 1\}^n \times \{0, 1\}^n$ is $\gamma$-subcube-like with
    respect to $(I, J)$ where $I, J \subseteq [n]$ if $\bm{x} \sim X$ is $(I, \gamma)$-structured and
    $\bm{y} \sim Y$ is $(J, \gamma)$-structured. In which case, we use $\fix(X) \coloneqq I$ and $\fix(Y)
    \coloneqq J$ to denote the fixed part of $X$ and $Y$ respectively.
\end{definition}

\begin{definition}[Subcube-like protocols~\cite{QuantumCommunicationTFNP}]
    A communication protocol $\Pi\colon \ZO^n \times \ZO^n \to S$ is $\gamma$-subcube-like if for every
    node $v \in \mathcal{N}(\Pi)$ in the protocol tree, $R_v$ is $\gamma$-subcube-like.
\end{definition}

\begin{definition}[Codimension]
    The codimension of a subcube-like rectangle $R = X \times Y$ is defined as the total number of fixed
    positions in $X$ and $Y$, denoted $\codim(R) \coloneqq |\fix(X)| + |\fix(Y)|$. 
    The codimension of a subcube-like protocol $\Pi$ is the maximum codimension of subcube-like
    rectangles associated with any nodes in the protocol tree of $\Pi$, denoted $\codim(\Pi)\coloneqq
    \max_{v\in \mathcal{N}(\Pi)} \codim(R_v)$. 
    
    
\end{definition}

\begin{lemma}[Density Restoring Partition~\cite{BPPlifting}]\label{lemma:density_restoring_partition}
    Let $\bm{x} \in \{0, 1\}^n$ be a random boolean vector with support $X \subseteq \ZO^n$ and $0 <
    \gamma < 1$ be a fixed parameter. There exists a partition
    $$
        X = X^1 \sqcup X^2 \sqcup X^3 \cdots \sqcup X^{r} 
    $$
    such that for each $j \in [r]$, $\bm{x} \mid \bm{x} \in X^j$ is $(I^j, \gamma)$-structured with
    respect to some $I^j \subseteq [n]$.

    Moreover, if we denote $p^{\ge j} \coloneqq \Pr[\bm{x} \in X^j \sqcup \cdots \sqcup X^{\ell} ]$, then it
    holds that:
    $$
        \Hinf(\bm{x}_{[m]\setminus I^j} \mid \bm{x} \in X^j) \ge \Hinf(\bm{x}) - \gamma |I^j| - \log (1/p^{\ge j}).
    $$
\end{lemma}

\subsection{Subcube-like protocols from general protocols}
\label{sec:subcube_from_general}
Göös et al.~\cite{QuantumCommunicationTFNP} show how to convert an arbitrary communication protocol into a subcube-like one.
Specifically, they prove the following.
\begin{lemma}[\cite{QuantumCommunicationTFNP}]
    Let $\Pi$ be a communication protocol of depth $d$ and $\epsilon > 0$. There exists a subcube-like
    protocol $\tilde{\Pi}$ of codimension $\codim(\tilde{\Pi}) = O(d / \varepsilon)$ such that
    \[
        \Pr_{\bm{x}, \bm{y}}[\Pi(\bm{x}, \bm{y}) \ne \tilde{\Pi}(\bm{x}, \bm{y})] \le \varepsilon.
    \]
\end{lemma}
Their bound is tight in the constant-error regime. However, it degenerates when $\varepsilon = O(d / n)$.

In this subsection, we give a more refined analysis of the reduction in~\cite{QuantumCommunicationTFNP},
which makes the bound applicable in the inverse polynomial error regime (when $d = \Omega(\log n)$).
We remark that such an analysis has been implicitly provided in~\cite{BPPlifting}.
\begin{lemma}
    \label{lemma:general_to_subcube}
    Let $\Pi$ be a communication protocol of depth $d$. There exists a $\gamma$-subcube-like protocol
    $\tilde{\Pi}$ of codimension $\codim(\tilde{\Pi}) = \frac{7}{1 - \gamma} \cdot d$ such that
    \[
        \Pr_{\bm{x}, \bm{y}}[\Pi(\bm{x}, \bm{y}) \ne \tilde{\Pi}(\bm{x}, \bm{y})] \le \exp(-d).
    \]
\end{lemma}

We include a simplified version of the algorithm for such conversion from \cite{QuantumCommunicationTFNP}
for completeness. This algorithm simulates a subcube-like protocol $\Pi'$ on an input $(x, y)$, given a
general protocol $\Pi$. 

\setcounter{algorithm}{1}
\begin{algorithm}
    \caption{\textbf{(simplified) conversion from \cite{QuantumCommunicationTFNP}}}\label{alg:subcube}
    \begin{algorithmic}
    \State $v \gets \text{root of } \Pi$, $X \times Y = \{0, 1\}^n \times \{0, 1\}^n$, $I \gets \varnothing$. 
    \While{$v \text{ is not a leaf}$}
        \State $v_0, v_1 \gets \text{ children of } v$
        \State \textbf{Suppose} Alice \textbf{sends} a bit at $v$ (otherwise swap $X$ and $Y$, $I$ and $J$)
        \State \textbf{Let} $X = X^0 \sqcup X^1$ be the partition according to the bit Alice sends
        \State \textbf{Let} $X^b = \bigsqcup_{i}X^{b,i}$ be the density-restoring partition (with
            parameter $\gamma$ and sets $I^i$, respectively), where $x\in X^b$.
        \State $X \gets X^{b,i}, I \gets I \cup I^i$ where $x \in X^i$
        \State Alice \textbf{sends} $(b, C(i))$ to Bob (here $C(i)$ is any encoding of $i$)
        \State $v \gets v^b$
    \EndWhile
    \State {\bf Output} the label $\Pi(v)$
    \end{algorithmic}
\end{algorithm}

\begin{proof}
    Let $\Pi'$ be as given by Algorithm 2 in~\cite{QuantumCommunicationTFNP}. More precisely, let the
    protocol tree of $\Pi'$ consist all the possible configurations at the end of each iteration plus the
    initial one as the root (so the tree is not necessarily binary). Observe that $\Pi'$ has depth $d$,
    though may have much larger communication complexity.
    
    For any $x, y \in \ZO^n$, we have $\Pi(x, y) = \Pi'(x, y)$. It suffices to show
    $\Pr_{\bm{x}, \bm{y}}[\codim(R(\bm{x}, \bm{y})) > \frac{7}{1 - \gamma} \cdot d] \le \exp(-d)$, where
    $R(x, y)$ is the unique rectangle associated with the leaves of $\Pi$ that contains $(x, y)$. The
    desired $\tilde{\Pi}$ can then be obtained by shaving all the nodes in the protocol tree of $\Pi'$
    associated with a rectangle of codimension greater than $\frac{7}{1 - \gamma} \cdot d$. 

    For each node $v\in \mathcal{N}(\Pi')$, define the entropy deficiency of $v$ as
    \[
        \Dinf(v)\coloneqq \Dinf(X_v)+\Dinf(Y_v),
    \]
    where
    \[
        \Dinf(X_v) \coloneqq
        n - |\fix(X_v)| - \Hinf(\bm{x}_{[n] \setminus \fix(X_v)}),\quad\quad \bm{x} \sim X_v
    \]
    and $\Dinf(Y_v)$ is defined analogously.
    
    Now consider running $\Pi'$ on $x,y$, and let $v_0,\ldots,v_{d}\in \mathcal{N}(\Pi')$ denote all the
    nodes on the execution path. Fix any $k\in [d]$ and let us simply use $u$ and $v$ to denote
    $v_{k - 1}$ and $v_k$. Suppose without loss of generality that it is Alice who sends a bit to Bob in
    the $k$-th iteration. Recall that in each iteration Alice first partitions $X_u=X_u^0\sqcup X_u^1$
    according to the bit she sends.
    Then she performs the density-restoring partition with parameter $\gamma$ on $X_u^b = X_u^{b, 1}
    \sqcup \ldots \sqcup X_u^{b, r}$ where $x\in X_u^b$. Finally, she determines the unique $X^{b, i}_u$
    that contains $x$. Then for the next configuration, $X_v = X^{b, i}_u$. Let us define
    \begin{flalign*}
      &q_u^b\coloneqq \Pr[\bm{x}\in X^b_u\mid \bm{x}\in X_u],\\
      &p_u^{b,\ge j}\coloneqq \Pr[\bm{x}\in \bigcup_{k\ge j} X^{b,k}_u\space \bigg| \space  \bm{x}\in
        X^b_u] \quad \forall j\in [r],\\
      &h_k(x,y)\coloneqq \log(1/q^b_u)+\log(1/p_u^{b,\ge i}),\\
      &n_k(x,y)\coloneqq |\fix(X_v)\setminus \fix(X_u)|.
    \end{flalign*}

    We have the following simple fact.
    \begin{fact}
        \label{fc:deficiency}
        $\Dinf(v) \le \Dinf(u) - (1 - \gamma) n_k(x, y) + h_k(x, y)$.
    \end{fact}

    \begin{proof}
        For proof see appendix \ref{sec:fc-deficiency}.
    \end{proof}
    
    Together with the nonnegativity of $\Dinf$, we can bound the codimension of $R(x, y)$ by $h(x, y)
    \coloneqq \sum_{k = 1}^{d} h_k(x,y)$ up to a multiplicative factor.
    
    \begin{claim}\label{claim:codim_bound}
        For every $x,y\in \{0,1\}^n$, $\codim(R(x,y))\le \frac{1}{1-\gamma}\cdot h(x,y)$.
    \end{claim}
    
    \begin{proof}
        Consider the path in the tree leading to the leaf $R(x, y)$, this path being of length
        $d$. Summing up the inequalities from \cref{fc:deficiency} along that path, we get:
        \[
            \Dinf(v_d) - \Dinf(v_0) \le -(1-\gamma) \sum_{j = 1}^{d } n_j(x,y) + \sum_{j = 1}^{d} h_j(x,
            y)
        \]

        Since $\Dinf(v_d)$ is non-negative and $\Dinf(v_0) = 0$, it follows that:
        \[
            \codim(R(x, y)) = \sum_{j = 1}^{d} n_j(x,y) \le \frac{1}{1-\gamma} \sum_{j = 1}^{d } h_j(x,
            y) = \frac{1}{1-\gamma}\cdot h(x, y).\qedhere
        \]
    \end{proof}
    We also observe that $h_k(\bm{x},\bm{y})$ has an exponential tail for each $k\in [d]$, even conditioned on any node $v$ of depth $k-1$ being reached.
    \begin{claim}\label{claim:tail_bound}
    For every node $v\in \mathcal{N}(\Pi')$ of depth $0\le k<d$ and threshold $\gamma\ge 0$,
    \[
    \Pr_{\bm{x},\bm{y}}[h_{k+1}(\bm{x},\bm{y})\ge 1+\gamma\mid \bm{v}_k=v]\le 2^{-\gamma}.
    \]
    \end{claim}
    \begin{proof}
    Let $\bm{b},\bm{i}$ be as defined in the $(k+1)$-th iteration of \Cref{alg:subcube} given $\bm{x},\bm{y}$.   
    We have
     \begin{align*}
         &\Pr_{\bm{x},\bm{y}}[h_{k+1}(\bm{x},\bm{y})\ge 1+\gamma\mid \bm{v}_k=v] \\ 
         =&\sum_{b\in \ZO} \Pr[\bm{b}=b\mid \bm{v}_k=v]\cdot \Pr[\log(1/q^b_v)+\log(1/p_v^{b,\ge\bm{i}})\ge 1+\gamma\mid \bm{b}=b,\bm{v}_k = v] \\
         =& \sum_{b\in \ZO} q^b_v\cdot\Pr\left[q^b_v \cdot p^{b,\ge \bm{i}} \le 2^{-\gamma-1} \space \bigg| \space \bm{b}=b,\bm{v}_k = v\right] \\
         \le& \sum_{b\in \ZO} q^b_v\cdot \min\left\{1,2^{-\gamma-1}\cdot q^b_v\right\}\\
         \le& 2^{-\gamma},
     \end{align*}
    where in the second last inequality, we use the property that $\Pr[p_v^{b,\ge \bm{i}}\le t\mid \bm{b}=b,\bm{v}_k=v]\le t$ for all $t\in[0,1]$.
    \end{proof}

    Finally, we need the following adaptive version of Bernstein's inequality, whose proof can be found in~\Cref{sec:concentration_proof}. 
    \begin{restatable}{lemma}{LemmaConcentration}
    \label{lemma:concentration}
    Let $\bm{a}_1,\ldots,\bm{a}_n\in \mathbb{R}$ be a random sequence of reals and $\zeta>0$ be some fixed parameter.
    If for any $1\le k\le n$ and $a_1,\ldots,a_{k-1}\in \mathbb{R}$ such that $\Pr[\bm{a}_1=a_1,\ldots,\bm{a}_{k-1}=a_{k-1}]>0$,
    \[
        \Pr[\bm{a}_k\ge x\mid \bm{a}_1=a_1,\ldots,\bm{a}_{k-1}=a_{k-1}]\le \exp(-\zeta x),
    \]
    then
    \[
        \Pr\left[\sum_{i=1}^n \bm{a}_i\ge \frac{4}{\zeta}n\right]\le\exp(-n).
    \]
\end{restatable}
\medskip
We are now ready to bound the codimension of $R(\bm{x},\bm{y})$.
Let $(\bm{a}_k\coloneqq h_k(\bm{x},\bm{y})-1)_{k\in [d]}\in \mathbb{R}^d$ be a random sequence of reals.
    By~\Cref{claim:tail_bound}, $\bm{a}$ satisfies the condition in \Cref{lemma:concentration} with $\zeta=\ln 2$. Therefore,
    \[
        \Pr[h(\bm{x},\bm{y})\ge 7d]=\Pr[\sum_{i=1}^d \bm{a}_i\ge 6d]\le\exp(-d).
    \]
    Together with~\Cref{claim:codim_bound}, we conclude that
    \[
        \Pr[\codim(R(\bm{x},\bm{y}))\ge \frac{7}{1-\gamma}\cdot d]\le \Pr[h(\bm{x},\bm{y})\ge 7d]\le\exp(-d). \qedhere
    \]
 
\end{proof}

\subsection{Lower bound against subcube-like protocols}\label{sec:lower_bound_subcube}

The following lemma is implicit in \cite{Condensation}, we include its proof in \cref{sec:expander-proof}
for completeness. In fact 

\begin{lemma}
    \label{lem:expander-closure}
    Let $0 < \beta < \alpha < 1$ and let $\Pi\colon \ZO^n \times \ZO^n\to S$ be a subcube-like protocol
    of codimension $d \coloneqq \codim(\Pi)$ where $d \le (\alpha - \beta)^2 r \Delta / 4$,
    and $G_1 = ([m],[n],E_1), G_2 = ([m],[n],E_2)$ be two $(r, \Delta, \alpha\Delta)$-expanders.
    Then there exist families $\{\Cl^X(v)\}_{v \in \mathcal{N}(\Pi)},
    \{\Cl^Y(v)\}_{v \in \mathcal{N}(\Pi)}$ of subsets of $[m]$ such that the following conditions hold:
    \begin{enumerate}
        \item For every non-root $v\in \mathcal{N}(\Pi)$, let $u$ denote $v$'s parent. Then
            $\Cl^X(u) \subseteq \Cl^X(v)$ and $\Cl^Y(u) \subseteq \Cl^Y(v)$.
        \item For every $v\in \mathcal{N}(\Pi)$, $G_1 - \Cl^X(v) - N(\Cl^X(v)) - \fix(X_v)$ and $G_2 -
            \Cl^Y(v) - N(\Cl^Y(v)) - \fix(Y_v)$ are both $(r,\Delta, \beta\Delta)$-expanders.
        \item For every $v \in \mathcal{N}(\Pi)$, $|\Cl^X(v)|, |\Cl^Y(v)| \le \frac{1}{\alpha - \beta}
            \cdot \frac{d}{\Delta}$.
    \end{enumerate}
\end{lemma}

\begin{lemma}\label{lemma:subcube_protocol_error}
    As in \cref{thm:bipartite-main} let $G_1 \coloneqq ([m], [n], E_1), G_2 \coloneqq ([m],[n],E_2)$ be two $(r, \Delta, \alpha
    \Delta)$-expanders, and $X, Y \coloneqq \{0, 1\}^n$. For each $i \in [m]$, let $C_i$ be a disjunction of variables in $\{ x_j\mid j\in
    N_{G_1}(i)\} \cup \{ y_j\mid j\in N_{G_2}(i)\}$ with arbitrary signs.
    Let $\Pi\colon \ZO^n\times \ZO^n\to [m]$ be a $\gamma$-subcube-like 
    communication protocol of $d \coloneqq \codim(\Pi)$. If
    $d \le \alpha^2 r\Delta / 4$, then
    \[
        \Pr_{\bm{x},\bm{y}}[C_{\bm{i}}(\bm{x},\bm{y})=0\mid \bm{i}=\Pi(\bm{x},\bm{y})]\le O(2^{-\gamma\alpha\Delta/2}\cdot d).
    \] 
\end{lemma}

\begin{proof}
    We rephrase the success probability of $\Pi$ as follows:
    Sample a random leaf $\bm{\ell}$ of $\Pi$ with probability $|R_{\bm{\ell}}|/2^{2n}$.
    Then
    \begin{equation}\label{eqn:succ_prob}
        \Pr_{\bm{x},\bm{y}}[C_{\bm{i}}(\bm{x},\bm{y})=0\mid
        \bm{i}=\Pi(\bm{x},\bm{y})]=\Exp_{\bm{\ell}}\Big[\Pr_{(\bm{x},\bm{y})\sim R_{\bm{\ell}}}[C_{\Pi(\bm{\ell})}(\bm{x},\bm{y})=0]\Big].
    \end{equation}

    Let $\{\Cl^X(v)\}_{v\in\mathcal{N}(\Pi)},\{\Cl^Y(v)\}_{v\in \mathcal{N}(\Pi)}$ be given by~\Cref{lem:expander-closure} with respect to $\Pi,G_1,G_2$ and $\beta=\alpha/2$. 
    For each node $v\in \mathcal{N}(\Pi)$, define $J_v\coloneqq
    \Cl^X(v)\cup \Cl^Y(v)$.
    We first observe that for each leaf $\ell$, $\Pi$ has low success probability on $R_\ell$ if
    $\Pi(\ell)\notin J_\ell$. 
    \begin{claim}
    \label{cl:not-in-intersection}
        Let $\ell$ be any leaf in the protocol tree of $\Pi$. Suppose that $i \not\in J_\ell$. Then
        \[\Pr_{(\bm{x},\bm{y})\sim R_{\ell}}[C_i(\bm{x},\bm{y})=0]\le 2^{-\gamma\alpha\Delta/2}.\]
    \end{claim}
    \begin{proof}
        By the definition of $J_\ell$, we have $i \not\in \Cl^X(\ell)$.
        Let $A \subseteq [n] \setminus (\fix(X) \cup N(\Cl^X(\ell)))$ be the set of neighbors of $i$ in $G_1 - \Cl^X(\ell) - N(\Cl^X(\ell)) - \fix(X_\ell)$, by the expansion we get $|A| \ge \alpha \Delta/2$.
        Since $X_\ell \times Y_\ell$ is $\gamma$-subcube-like we have that $\bm{x}_{[n] \setminus \fix(X_\ell)}$ is $\gamma$-spread. In particular, $\Hinf(\bm{x}_A) \ge \gamma |A| \ge \gamma \alpha \Delta/2$.
        Let $\tau \in \{0,1\}^A$ be the unique assignment that violates all literals of $C_i$ in $A$. The min-entropy bound above then implies
        \(\Pr[C_i(\bm{x}, \bm{y}) = 0] \le \Pr[\bm{x}_A = \tau] \le 2^{-\gamma\alpha\Delta/2}.\)
    \end{proof}
On the other hand, unfortunately, it is possible that
\[
p_i(\ell)\coloneqq \Pr_{(\bm{x},\bm{y})\sim R_{\ell}}[C_i(\bm{x},\bm{y})=0]
\]
is close to $1$ for some $i\in J_\ell$. Nevertheless, we can show that this can happen only for a small fraction of leaves. 

\begin{claim}
    Let $\bm{\ell}$ be a random leaf sampled as stated above. Then
    \[
    \Exp_{\bm{\ell}}\left[\sum_{i\in J_{\bm{\ell}}} p_i(\bm{\ell})\right]\le 2^{-\gamma\alpha\Delta/2}\cdot d.
    \]
\end{claim}
\begin{proof}
First, we can write
\begin{align*}
     \Exp_{\bm{\ell}}\left[\sum_{i\in J_{\bm{\ell}}} p_i(\bm{\ell})\right] 
     &=\sum_{i \in [m]} \Exp_{\bm{\ell}}[\mathbf{1}_{i \in J_{\bm{\ell}}} \cdot p_i(\bm{\ell})]\\
    \text{(where $\ell_{\bm{x},\bm{y}}$ is the leaf containing $(\bm{x}, \bm{y})$)} &=\sum_{i \in [m]} \Pr_{\bm{x},\bm{y}}[i \in J_{\ell_{\bm{x},\bm{y}}} \,\land\, C_i(\bm{x}, \bm{y}) = 0]\\
    &=\sum_{i \in [m]} \Pr_{\bm{x},\bm{y}}[C_i(\bm{x},\bm{y})=0\mid i\in J_{\ell_{\bm{x},\bm{y}}}]\cdot \Pr_{\bm{x},\bm{y}}[i\in J_{\ell_{\bm{x},\bm{y}}}]\\
     &\le \max_{i \in [m]} \Pr_{\bm{x},\bm{y}}[C_i(\bm{x}, \bm{y})=0 \mid i \in J_{\ell_{\bm{x},\bm{y}}}] \cdot \sum_{i \in [m]} \Pr_{\bm{x},\bm{y}}[i\in J_{\ell_{\bm{x},\bm{y}}}]\\
     &=\max_{i \in [m]} \Pr_{\bm{x},\bm{y}}[C_i(\bm{x}, \bm{y})=0 \mid i \in J_{\ell_{\bm{x},\bm{y}}}] \cdot \Exp_{\bm{\ell}}[|J_{\bm{\ell}}|].
\end{align*}
Observe that $|J_{\ell}|\le d$ for every leaf $\ell$, it suffices to show
\[
\Pr_{\bm{x},\bm{y}}[C_i(\bm{x}, \bm{y})=0 \mid i \in J_{\ell_{\bm{x},\bm{y}}}]\le 2^{-\gamma\alpha\Delta/2}.
\]
for every $i \in [m]$.
Now let us fix an arbitrary $i \in [m]$. The event ``$i \in J_{\ell_{\bm{x},\bm{y}}}$'' can be reinterpreted as follows: with 
\[V_i \coloneqq \{v \in \mathcal{N}(\Pi) \mid i \in J_v \text{ and the parent of $v$ does not satisfy that}\}\]
we have that $i \in J_{\ell_{\bm{x},\bm{y}}}$ if and only if $(\bm{x}, \bm{y}) \in \bigsqcup_{v \in V_i} R_v$ (the rectangles form a partition since the nodes in $V_i$ are maximally close to the root). Then
\[\Pr_{\bm{x},\bm{y}}[C_i(\bm{x}, \bm{y})=0 \mid i \in J_{\ell_{\bm{x},\bm{y}}}] \le \sum_{v \in V_i} \Pr_{\bm{x},\bm{y}}[(\bm{x}, \bm{y}) \in R_v\mid i \in J_{\ell_{\bm{x},\bm{y}}}] \cdot \Pr_{\bm{x},\bm{y}}[C_i(\bm{x}, \bm{y}) = 0 \mid (\bm{x}, \bm{y}) \in R_v].\] 
Since the right-hand side is a convex combination of $\Pr[C_i(\bm{x},\bm{y}) = 0 \mid (\bm{x},\bm{y}) \in R_v]$ for $v \in V_i$, it suffices to bound the maximum of these probabilities. 

The crucial observation to conclude the proof is that $i \not\in \Cl^X(v)$ if Bob spoke in the parent node of $v$ and $i \not\in \Cl^Y(v)$ if Alice spoke in that node. 
In any case, an argument similar to that in \cref{cl:not-in-intersection} applies and we have $\Pr[C_i(\bm{x},\bm{y}) = 0 \mid (\bm{x},\bm{y}) \in R_v] \le 2^{-\gamma\alpha\Delta/2}$, which concludes the proof.
\end{proof}
Now we are ready to show the desired bound. Combining the above two claims, we have
\begin{align*}
\eqref{eqn:succ_prob}&=\Pr[\Pi(\bm{\ell})\notin J_{\bm{\ell}}]\cdot \Exp_{\bm{\ell}}[\Pr_{(\bm{x},\bm{y})\sim R_{\bm{\ell}}}[C_{\Pi(\bm{\ell})}(\bm{x},\bm{y})=0]\mid \Pi(\bm{\ell})\notin J_{\bm{\ell}} ]\\
&+\Pr[\Pi(\bm{\ell})\in J_{\bm{\ell}}]\cdot\Exp_{\bm{\ell}}[\Pr_{(\bm{x},\bm{y})\sim R_{\bm{\ell}}}[C_{\Pi(\bm{\ell})}(\bm{x},\bm{y})=0]\mid \Pi(\bm{\ell})\in J_{\bm{\ell}} ]\\
&\le 2^{-\gamma\alpha\Delta/2}+\Exp_{\bm{\ell}}[\sum_{i\in J_{\bm{\ell}}}p_i(\bm{\ell})]\\
&=O(d/2^{\gamma\alpha\Delta/2}).\qedhere\end{align*}
\end{proof}

\subsection{Proof of \Cref{thm:bipartite-main}}
\label{sec:proof_bipartite}

We first restate the theorem for convenience.
\TheoremMainBipartite*

\begin{proof}
    Let $\tilde{\Pi}$ be a subcube-like protocol given by~\Cref{lemma:general_to_subcube} with respect to
    $\Pi$ and $\gamma = \alpha$. Then $\codim(\Pi') \le \frac{7d}{1 - \alpha}$. Moreover,
    \[
        \Pr_{\bm{x}, \bm{y}}[\Pi(\bm{x}, \bm{y}) \ne \tilde{\Pi}(\bm{x}, \bm{y})] \le \exp(-d).
    \]
    We can then apply~\Cref{lemma:subcube_protocol_error} and conclude that
    \[
        \Pr_{\bm{x}, \bm{y}}[C_{\bm{i}}(\bm{x}, \bm{y}) = 0 \mid \bm{i} = \Pi(\bm{x}, \bm{y})] \le
        \Pr_{\bm{x}, \bm{y}}[C_{\bm{i}}(\bm{x}, \bm{y}) = 0 \mid
        \bm{i}=\tilde{\Pi}(\bm{x},\bm{y})] + \exp(-d) = d \cdot 2^{-\Omega(\Delta)} + \exp(-d).
        \qedhere
    \]
\end{proof}

\DeclareUrlCommand{\Doi}{\urlstyle{sf}}
\renewcommand{\path}[1]{\small\Doi{#1}}
\renewcommand{\url}[1]{\href{#1}{\small\Doi{#1}}}
\bibliographystyle{alphaurl}
\bibliography{references}

\newcommand{\etalchar}[1]{$^{#1}$}
\begin{thebibliography}{ABSRW04}

\bibitem[ABSRW04]{ABRW04}
Michael Alekhnovich, Eli Ben-Sasson, Alexander~A. Razborov, and Avi Wigderson.
\newblock Pseudorandom generators in propositional proof complexity.
\newblock {\em {SIAM} J. Comput.}, 34(1):67--88, 2004.
\newblock \href {https://doi.org/10.1137/S0097539701389944} {\path{doi:10.1137/S0097539701389944}}.

\bibitem[Ale11]{Alekhnovich11}
Michael Alekhnovich.
\newblock Lower bounds for k-dnf resolution on random 3-cnfs.
\newblock {\em Comput. Complex.}, 20(4):597--614, 2011.
\newblock \href {https://doi.org/10.1007/s00037-011-0026-0} {\path{doi:10.1007/s00037-011-0026-0}}.

\bibitem[AR03]{AR03}
Michael Alekhnovich and Alexander~A. Razborov.
\newblock Lower bounds for polynomial calculus: {N}on-binomial case.
\newblock {\em Proceedings of the Steklov Institute of Mathematics}, 242:18--35, 2003.
\newblock Available at \url{http://people.cs.uchicago.edu/~razborov/files/misha.pdf}. Preliminary version in \emph{FOCS~'01}.

\bibitem[BKPS02]{BKPS02}
Paul Beame, Richard~M. Karp, Toniann Pitassi, and Michael~E. Saks.
\newblock The efficiency of resolution and davis--putnam procedures.
\newblock {\em {SIAM} J. Comput.}, 31(4):1048--1075, 2002.
\newblock \href {https://doi.org/10.1137/S0097539700369156} {\path{doi:10.1137/S0097539700369156}}.

\bibitem[BPS07]{BPS07}
Paul Beame, Toniann Pitassi, and Nathan Segerlind.
\newblock Lower bounds for lov[a-acute]sz--schrijver systems and beyond follow from multiparty communication complexity.
\newblock {\em {SIAM} J. Comput.}, 37(3):845--869, 2007.
\newblock \href {https://doi.org/10.1137/060654645} {\path{doi:10.1137/060654645}}.

\bibitem[BW25]{BW24}
Paul Beame and Michael Whitmeyer.
\newblock {Multiparty Communication Complexity of Collision-Finding and Cutting Planes Proofs of Concise Pigeonhole Principles}.
\newblock In Keren Censor-Hillel, Fabrizio Grandoni, Jo\"{e}l Ouaknine, and Gabriele Puppis, editors, {\em 52nd International Colloquium on Automata, Languages, and Programming (ICALP 2025)}, volume 334 of {\em Leibniz International Proceedings in Informatics (LIPIcs)}, pages 21:1--21:20, Dagstuhl, Germany, 2025. Schloss Dagstuhl -- Leibniz-Zentrum f{\"u}r Informatik.
\newblock URL: \url{https://drops.dagstuhl.de/entities/document/10.4230/LIPIcs.ICALP.2025.21}, \href {https://doi.org/10.4230/LIPIcs.ICALP.2025.21} {\path{doi:10.4230/LIPIcs.ICALP.2025.21}}.

\bibitem[CS88]{ChvSem88}
Va\v{s}ek Chv\'{a}tal and Endre Szemer{\'e}di.
\newblock Many hard examples for resolution.
\newblock {\em J. ACM}, 35(4):759--768, October 1988.
\newblock URL: \url{http://doi.acm.org/10.1145/48014.48016}, \href {https://doi.org/10.1145/48014.48016} {\path{doi:10.1145/48014.48016}}.

\bibitem[dRFJ{\etalchar{+}}25]{dRFJNP24}
Susanna~F. de~Rezende, Noah Fleming, Duri~Andrea Janett, Jakob Nordstr\"{o}m, and Shuo Pang.
\newblock Truly supercritical trade-offs for resolution, cutting planes, monotone circuits, and weisfeiler–leman.
\newblock In {\em Proceedings of the 57th Annual ACM Symposium on Theory of Computing}, STOC '25, page 1371–1382, New York, NY, USA, 2025. Association for Computing Machinery.
\newblock \href {https://doi.org/10.1145/3717823.3718271} {\path{doi:10.1145/3717823.3718271}}.

\bibitem[dRNV16]{RNV16}
Susanna~F. de~Rezende, Jakob Nordstr{\"{o}}m, and Marc Vinyals.
\newblock How limited interaction hinders real communication (and what it means for proof and circuit complexity).
\newblock In Irit Dinur, editor, {\em {IEEE} 57th Annual Symposium on Foundations of Computer Science, {FOCS} 2016, 9-11 October 2016, Hyatt Regency, New Brunswick, New Jersey, {USA}}, pages 295--304. {IEEE} Computer Society, 2016.
\newblock \href {https://doi.org/10.1109/FOCS.2016.40} {\path{doi:10.1109/FOCS.2016.40}}.

\bibitem[DSS16]{HardnessLearningDNFs}
Amit Daniely and Shai Shalev-Shwartz.
\newblock Complexity theoretic limitations on learning dnf's.
\newblock In Vitaly Feldman, Alexander Rakhlin, and Ohad Shamir, editors, {\em 29th Annual Conference on Learning Theory}, volume~49 of {\em Proceedings of Machine Learning Research}, pages 815--830, Columbia University, New York, New York, USA, 23--26 Jun 2016. PMLR.
\newblock URL: \url{https://proceedings.mlr.press/v49/daniely16.html}.

\bibitem[Fei02]{Feige02}
Uriel Feige.
\newblock Relations between average case complexity and approximation complexity.
\newblock In {\em Proceedings of the 17th Annual {IEEE} Conference on Computational Complexity, Montr{\'{e}}al, Qu{\'{e}}bec, Canada, May 21-24, 2002}, page~5. {IEEE} Computer Society, 2002.
\newblock URL: \url{http://doi.ieeecomputersociety.org/10.1109/CCC.2002.10006}, \href {https://doi.org/10.1109/CCC.2002.10006} {\path{doi:10.1109/CCC.2002.10006}}.

\bibitem[FPPR22]{FPPR22}
Noah Fleming, Denis Pankratov, Toniann Pitassi, and Robert Robere.
\newblock {Random $\Theta(\log n)$-CNFs are Hard for Cutting Planes}.
\newblock {\em J. {ACM}}, 69(3):19:1--19:32, 2022.
\newblock \href {https://doi.org/10.1145/3486680} {\path{doi:10.1145/3486680}}.

\bibitem[GGJL25]{QuantumCommunicationTFNP}
Mika G\"{o}\"{o}s, Tom Gur, Siddhartha Jain, and Jiawei Li.
\newblock Quantum communication advantage in tfnp.
\newblock In {\em Proceedings of the 57th Annual ACM Symposium on Theory of Computing}, STOC '25, page 1465–1475, New York, NY, USA, 2025. Association for Computing Machinery.
\newblock \href {https://doi.org/10.1145/3717823.3718155} {\path{doi:10.1145/3717823.3718155}}.

\bibitem[GGKS20]{GGKS20}
Ankit Garg, Mika G{\"{o}}{\"{o}}s, Pritish Kamath, and Dmitry Sokolov.
\newblock Monotone circuit lower bounds from resolution.
\newblock {\em Theory Comput.}, 16:1--30, 2020.
\newblock URL: \url{https://doi.org/10.4086/toc.2020.v016a013}, \href {https://doi.org/10.4086/TOC.2020.V016A013} {\path{doi:10.4086/TOC.2020.V016A013}}.

\bibitem[GJPW18]{GJPW18}
Mika G{\"{o}}{\"{o}}s, T.~S. Jayram, Toniann Pitassi, and Thomas Watson.
\newblock Randomized communication versus partition number.
\newblock {\em {ACM} Trans. Comput. Theory}, 10(1):4:1--4:20, 2018.
\newblock \href {https://doi.org/10.1145/3170711} {\path{doi:10.1145/3170711}}.

\bibitem[GJW18]{GJW18}
Mika G{\"{o}}{\"{o}}s, Rahul Jain, and Thomas Watson.
\newblock Extension complexity of independent set polytopes.
\newblock {\em {SIAM} J. Comput.}, 47(1):241--269, 2018.
\newblock \href {https://doi.org/10.1137/16M109884X} {\path{doi:10.1137/16M109884X}}.

\bibitem[GKRS19]{GKRS19}
Mika G{\"{o}}{\"{o}}s, Pritish Kamath, Robert Robere, and Dmitry Sokolov.
\newblock Adventures in monotone complexity and {TFNP}.
\newblock In Avrim Blum, editor, {\em 10th Innovations in Theoretical Computer Science Conference, {ITCS} 2019, January 10-12, 2019, San Diego, California, {USA}}, volume 124 of {\em LIPIcs}, pages 38:1--38:19. Schloss Dagstuhl - Leibniz-Zentrum f{\"{u}}r Informatik, 2019.
\newblock URL: \url{https://doi.org/10.4230/LIPIcs.ITCS.2019.38}, \href {https://doi.org/10.4230/LIPICS.ITCS.2019.38} {\path{doi:10.4230/LIPICS.ITCS.2019.38}}.

\bibitem[GLM{\etalchar{+}}16]{GLMWZ16}
Mika G{\"{o}}{\"{o}}s, Shachar Lovett, Raghu Meka, Thomas Watson, and David Zuckerman.
\newblock Rectangles are nonnegative juntas.
\newblock {\em {SIAM} J. Comput.}, 45(5):1835--1869, 2016.
\newblock \href {https://doi.org/10.1137/15M103145X} {\path{doi:10.1137/15M103145X}}.

\bibitem[GMRS25]{MGKS25}
Mika G\"{o}\"{o}s, Gilbert Maystre, Kilian Risse, and Dmitry Sokolov.
\newblock Supercritical tradeoffs for monotone circuits.
\newblock In {\em Proceedings of the 57th Annual ACM Symposium on Theory of Computing}, STOC '25, page 1359–1370, New York, NY, USA, 2025. Association for Computing Machinery.
\newblock \href {https://doi.org/10.1145/3717823.3718229} {\path{doi:10.1145/3717823.3718229}}.

\bibitem[GNRS24]{Condensation}
Mika G{\"{o}}{\"{o}}s, Ilan Newman, Artur Riazanov, and Dmitry Sokolov.
\newblock Hardness condensation by restriction.
\newblock In Bojan Mohar, Igor Shinkar, and Ryan O'Donnell, editors, {\em Proceedings of the 56th Annual {ACM} Symposium on Theory of Computing, {STOC} 2024, Vancouver, BC, Canada, June 24-28, 2024}, pages 2016--2027. {ACM}, 2024.
\newblock \href {https://doi.org/10.1145/3618260.3649711} {\path{doi:10.1145/3618260.3649711}}.

\bibitem[GP18a]{GP18-cbs}
Mika G{\"{o}}{\"{o}}s and Toniann Pitassi.
\newblock Communication lower bounds via critical block sensitivity.
\newblock {\em {SIAM} J. Comput.}, 47(5):1778--1806, 2018.
\newblock \href {https://doi.org/10.1137/16M1082007} {\path{doi:10.1137/16M1082007}}.

\bibitem[GP18b]{GP18}
Mika G{\"{o}}{\"{o}}s and Toniann Pitassi.
\newblock Communication lower bounds via critical block sensitivity.
\newblock {\em {SIAM} Journal on Computing}, 47(5):1778--1806, 2018.
\newblock \href {https://doi.org/10.1137/16M1082007} {\path{doi:10.1137/16M1082007}}.

\bibitem[GPW20]{BPPlifting}
Mika G{\"{o}}{\"{o}}s, Toniann Pitassi, and Thomas Watson.
\newblock Query-to-communication lifting for {BPP}.
\newblock {\em {SIAM} Journal on Computing}, 49(4), 2020.
\newblock \href {https://doi.org/10.1137/17M115339X} {\path{doi:10.1137/17M115339X}}.

\bibitem[Gri01]{Grig01}
Dima Grigoriev.
\newblock Linear lower bound on degrees of positivstellensatz calculus proofs for the parity.
\newblock {\em Theoretical Computer Science}, 259(1):613--622, 2001.
\newblock URL: \url{http://www.sciencedirect.com/science/article/pii/S0304397500001572}, \href {https://doi.org/10.1016/S0304-3975(00)00157-2} {\path{doi:10.1016/S0304-3975(00)00157-2}}.

\bibitem[HN12]{HN12}
Trinh Huynh and Jakob Nordstr{\"{o}}m.
\newblock On the virtue of succinct proofs: amplifying communication complexity hardness to time-space trade-offs in proof complexity.
\newblock In Howard~J. Karloff and Toniann Pitassi, editors, {\em Proceedings of the 44th Symposium on Theory of Computing Conference, {STOC} 2012, New York, NY, USA, May 19 - 22, 2012}, pages 233--248. {ACM}, 2012.
\newblock \href {https://doi.org/10.1145/2213977.2214000} {\path{doi:10.1145/2213977.2214000}}.

\bibitem[HP17]{HrubesP17}
Pavel Hrubes and Pavel Pudl{\'{a}}k.
\newblock Random formulas, monotone circuits, and interpolation.
\newblock In Chris Umans, editor, {\em 58th {IEEE} Annual Symposium on Foundations of Computer Science, {FOCS} 2017, Berkeley, CA, USA, October 15-17, 2017}, pages 121--131. {IEEE} Computer Society, 2017.
\newblock \href {https://doi.org/10.1109/FOCS.2017.20} {\path{doi:10.1109/FOCS.2017.20}}.

\bibitem[IPU94]{IPU94}
Russell Impagliazzo, Toniann Pitassi, and Alasdair Urquhart.
\newblock Upper and lower bounds for tree-like cutting planes proofs.
\newblock In {\em Proceedings of the Ninth Annual Symposium on Logic in Computer Science {(LICS} '94), Paris, France, July 4-7, 1994}, pages 220--228. {IEEE} Computer Society, 1994.
\newblock \href {https://doi.org/10.1109/LICS.1994.316069} {\path{doi:10.1109/LICS.1994.316069}}.

\bibitem[IR21]{ItsyksonR21}
Dmitry Itsykson and Artur Riazanov.
\newblock Proof complexity of natural formulas via communication arguments.
\newblock In Valentine Kabanets, editor, {\em 36th Computational Complexity Conference, {CCC} 2021, July 20-23, 2021, Toronto, Ontario, Canada (Virtual Conference)}, volume 200 of {\em LIPIcs}, pages 3:1--3:34. Schloss Dagstuhl - Leibniz-Zentrum f{\"{u}}r Informatik, 2021.
\newblock URL: \url{https://doi.org/10.4230/LIPIcs.CCC.2021.3}, \href {https://doi.org/10.4230/LIPICS.CCC.2021.3} {\path{doi:10.4230/LIPICS.CCC.2021.3}}.

\bibitem[IS20]{IS20}
Dmitry Itsykson and Dmitry Sokolov.
\newblock Resolution over linear equations modulo two.
\newblock {\em Ann. Pure Appl. Log.}, 171(1), 2020.
\newblock URL: \url{https://doi.org/10.1016/j.apal.2019.102722}, \href {https://doi.org/10.1016/J.APAL.2019.102722} {\path{doi:10.1016/J.APAL.2019.102722}}.

\bibitem[Kra97]{Krajicek97}
Jan Kraj{\'{\i}}cek.
\newblock Interpolation theorems, lower bounds for proof systems, and independence results for bounded arithmetic.
\newblock {\em J. Symb. Log.}, 62(2):457--486, 1997.
\newblock \href {https://doi.org/10.2307/2275541} {\path{doi:10.2307/2275541}}.

\bibitem[KW90]{KW90}
Mauricio Karchmer and Avi Wigderson.
\newblock Monotone circuits for connectivity require super-logarithmic depth.
\newblock {\em {SIAM} J. Discret. Math.}, 3(2):255--265, 1990.
\newblock \href {https://doi.org/10.1137/0403021} {\path{doi:10.1137/0403021}}.

\bibitem[LMM{\etalchar{+}}22]{LMMPZ22}
Shachar Lovett, Raghu Meka, Ian Mertz, Toniann Pitassi, and Jiapeng Zhang.
\newblock Lifting with sunflowers.
\newblock In Mark Braverman, editor, {\em 13th Innovations in Theoretical Computer Science Conference, {ITCS} 2022, January 31 - February 3, 2022, Berkeley, CA, {USA}}, volume 215 of {\em LIPIcs}, pages 104:1--104:24. Schloss Dagstuhl - Leibniz-Zentrum f{\"{u}}r Informatik, 2022.
\newblock URL: \url{https://doi.org/10.4230/LIPIcs.ITCS.2022.104}, \href {https://doi.org/10.4230/LIPICS.ITCS.2022.104} {\path{doi:10.4230/LIPICS.ITCS.2022.104}}.

\bibitem[LNNW95]{LNNW95}
L{\'{a}}szl{\'{o}} Lov{\'{a}}sz, Moni Naor, Ilan Newman, and Avi Wigderson.
\newblock Search problems in the decision tree model.
\newblock {\em {SIAM} J. Discret. Math.}, 8(1):119--132, 1995.
\newblock \href {https://doi.org/10.1137/S0895480192233867} {\path{doi:10.1137/S0895480192233867}}.

\bibitem[MPZ02]{MezParZec02}
Marc Mezard, Giorgio Parisi, and Riccardo Zecchina.
\newblock Analytic and algorithmic solution of random satisfiability problems.
\newblock {\em Science (New York, N.Y.)}, 297:812--815, 09 2002.
\newblock \href {https://doi.org/10.1126/science.1073287} {\path{doi:10.1126/science.1073287}}.

\bibitem[PR18]{PR18-null}
Toniann Pitassi and Robert Robere.
\newblock Lifting nullstellensatz to monotone span programs over any field.
\newblock In Ilias Diakonikolas, David Kempe, and Monika Henzinger, editors, {\em Proceedings of the 50th Annual {ACM} {SIGACT} Symposium on Theory of Computing, {STOC} 2018, Los Angeles, CA, USA, June 25-29, 2018}, pages 1207--1219. {ACM}, 2018.
\newblock \href {https://doi.org/10.1145/3188745.3188914} {\path{doi:10.1145/3188745.3188914}}.

\bibitem[Pud97]{Pudlak97}
Pavel Pudl{\'{a}}k.
\newblock Lower bounds for resolution and cutting plane proofs and monotone computations.
\newblock {\em J. Symb. Log.}, 62(3):981--998, 1997.
\newblock \href {https://doi.org/10.2307/2275583} {\path{doi:10.2307/2275583}}.

\bibitem[Raz90]{Razborov90}
Alexander~A. Razborov.
\newblock Applications of matrix methods to the theory of lower bounds in computational complexity.
\newblock {\em Comb.}, 10(1):81--93, 1990.
\newblock \href {https://doi.org/10.1007/BF02122698} {\path{doi:10.1007/BF02122698}}.

\bibitem[RPRC16]{RPRC16}
Robert Robere, Toniann Pitassi, Benjamin Rossman, and Stephen~A. Cook.
\newblock Exponential lower bounds for monotone span programs.
\newblock In Irit Dinur, editor, {\em {IEEE} 57th Annual Symposium on Foundations of Computer Science, {FOCS} 2016, 9-11 October 2016, Hyatt Regency, New Brunswick, New Jersey, {USA}}, pages 406--415. {IEEE} Computer Society, 2016.
\newblock \href {https://doi.org/10.1109/FOCS.2016.51} {\path{doi:10.1109/FOCS.2016.51}}.

\bibitem[Sok20]{S20}
Dmitry Sokolov.
\newblock (semi)algebraic proofs over {\(\pm\)}1 variables.
\newblock In Konstantin Makarychev, Yury Makarychev, Madhur Tulsiani, Gautam Kamath, and Julia Chuzhoy, editors, {\em Proceedings of the 52nd Annual {ACM} {SIGACT} Symposium on Theory of Computing, {STOC} 2020, Chicago, IL, USA, June 22-26, 2020}, pages 78--90. {ACM}, 2020.
\newblock \href {https://doi.org/10.1145/3357713.3384288} {\path{doi:10.1145/3357713.3384288}}.

\bibitem[Sok24]{S24}
Dmitry Sokolov.
\newblock {Random $(\log n)$-CNF Are Hard for Cutting Planes (Again)}.
\newblock In Bojan Mohar, Igor Shinkar, and Ryan O'Donnell, editors, {\em Proceedings of the 56th Annual {ACM} Symposium on Theory of Computing, {STOC} 2024, Vancouver, BC, Canada, June 24-28, 2024}, pages 2008--2015. {ACM}, 2024.
\newblock \href {https://doi.org/10.1145/3618260.3649636} {\path{doi:10.1145/3618260.3649636}}.

\bibitem[SS22]{SofronovaS22}
Anastasia Sofronova and Dmitry Sokolov.
\newblock A lower bound for \emph{k}-dnf resolution on random {CNF} formulas via expansion.
\newblock {\em Electron. Colloquium Comput. Complex.}, {TR22-054}, 2022.
\newblock URL: \url{https://eccc.weizmann.ac.il/report/2022/054}.

\bibitem[WYZ23]{WYZ23}
Shuo Wang, Guangxu Yang, and Jiapeng Zhang.
\newblock {Communication Complexity of Set-Intersection Problems and Its Applications}.
\newblock Technical report, ECCC, 2023.
\newblock URL: \url{https://eccc.weizmann.ac.il/report/2023/164}.

\bibitem[YZ24]{YZ24}
Guangxu Yang and Jiapeng Zhang.
\newblock Communication lower bounds for collision problems via density increment arguments.
\newblock In {\em Proceedings of the 56th Annual ACM Symposium on Theory of Computing}, STOC 2024, page 630–639, New York, NY, USA, 2024. Association for Computing Machinery.
\newblock \href {https://doi.org/10.1145/3618260.3649607} {\path{doi:10.1145/3618260.3649607}}.

\end{thebibliography}
\appendix

\section{Proof of \cref{lem:partition}}
\label{sec:proof-of-partititon}
 We start by proving \cref{item:whp-satA} (\cref{item:whp-satB} is analogous). 
    Let $N^{\bm{X}}(i)$ for $i \in [m]$ be the set of neighbors of $i$ in $\bm{X}$ and $N^{\bm{Y}}(i)$
    --- in $\bm{Y}$. Then $\Err_{\bm{X}} = \{i \in [m] \mid |N^{\bm{X}}(i)| > (1-\delta)\Delta\}$.
    We then write
    \begin{align*}
      \Exp[|\Err_{\bm{X}}|] &= \sum_{i \in [m]} \Pr\big[|N^{\bm{X}}(i)| > (1-\delta)\Delta\big] \\
                            &=\sum_{i \in [m]}\sum_{S \subseteq N(i)\colon |S|\ge(1-\delta)\Delta}
                              \Pr[\bm{X} \cup N(i) = S]\\
                            &=m \sum_{j \le \delta \Delta} \binom{m}{m-j} 2^{-\Delta}\\
                            &\le m 2^{-(1-\H(\delta))\Delta}\\
                            &=\alpha n \cdot 2^{\H(\delta)\Delta}\\
                            &=\alpha n^{1+c \H(\delta)}
    \end{align*}
    
    On the other hand for every $i \in \Err_{\bm{X}}$ we have
    \[
        \Pr_{\bm{x} \sim \{0 ,1\}^{\bm{X}}}[C_i(\bm{x},\cdot) \not\equiv 1] = 2^{-|N^{\bm{X}}(i)|} \le
        2^{-(1 - \delta) \Delta} = n^{-c(1 - \delta)}.
    \] 
    Then by a union bound we get
    \[
        \Pr_{\bm{x} \sim \{0,1\}^{\bm{X}}}[\exists i \in \Err_{\bm{X}}\colon C_i(\bm{x},\cdot)\not\equiv
        1] \le |\Err_{\bm{X}}| \cdot n^{-c(1-\delta)}.
    \]
    Then by Markov's inequality applied to $|\Err_{\bm{X}}|$ with probability $1-\epsilon$ over $\bm{X}$
    we have

    \begin{align*}
      \Pr_{\bm{x} \sim \{0,1\}^{\bm{X}}}[\exists i \in \Err_{\bm{X}}\colon C_i(\bm{x},\cdot)\not\equiv 1]
      &\le 1/\epsilon \cdot \Exp[|\Err_{\bm{X}}|] \cdot n^{-c (1-\delta)}\\
      &=\alpha/\epsilon \cdot n^{1 + c \H(\delta) - c(1-\delta)}\\
      &=\alpha/\epsilon \cdot n^{1 - c(1-\delta-\H(\delta))}
    \end{align*}

    Now it remains to prove \cref{item:whp-expander}. First, let $\bm{G} \coloneqq ([m], [n], E_{\bm{X}}
    \sqcup E_{\bm{Y}})$ be the union of $G_{\bm{X}}$ and $G_{\bm{Y}}$. By \cref{lm:exp-parameters}
    whp over $\bm{\varphi}$ the graph $\bm{G}$ is an $(r, \Delta, (1-\eta)\Delta)$-expander for any $\eta$
    and $r = \Omega_{\eta}\left(\frac{n}{\Delta}\right)$. Now it is sufficient to show $G_{\bm{X}} -
    \Err_{\bm{Y}}$ is an $(r, \Delta, (\delta-2\eta)\Delta)$-expander whp, $G_{\bm{Y}} - \Err_{\bm{X}}$
    is distributed identically to $G_{\bm{X}}- \Err_{\bm{Y}}$ and removing additional nodes from the
    left-hand side does not reduce expansion. 

    We in fact show that conditioned on the fact that $\bm{G}$ is an $(r, \Delta,
    (1-\eta)\Delta)$-expander, $G_{\bm{X}} - \Err_{\bm{Y}}$ is $(r,\Delta,
    (\delta-2\eta)\Delta)$-expander with probability $1$. Consider an arbitrary subset $U \subseteq [m]$
    of size at most $r$. For every such subset we need to have $|N(U) \setminus \bm{Y} \setminus
    N(\Err_{\bm{Y}})| \ge (\delta -2\eta) \Delta |U \setminus \Err_{\bm{Y}}|$. Here and below $N(S) =
    N_{\bm{G}}(S)$. Consider the set $\partial U \coloneqq \{v \in N(U) \mid v \text{ is connected with a
        single node in } U\}$. Then $|\partial U| \ge (1-2\eta)|U|\Delta$: indeed the number of edges
    incident to $U$ can be estimated in two ways: 
    \[
        \Delta |U| = |E \cap (U \times [n])| \ge |\partial U| + 2(|N(U)|-|\partial U|) \ge
        2(1-\eta)\Delta|U| - |\partial U|.
    \] 

    Then we can partition the set $N(U)$ into sets $N_i$ for $i \in U$ where $N_i \subseteq N(i)$ and
    $|N_i| \ge (1-2\eta)\Delta$: find a node $i \in U$ such that $|\partial U \cap N(i)| \ge
    (1-2\eta)\Delta$, let $N_i \coloneqq \partial U \cap N(i)$ and continue the process for $U \setminus
    \{i\}$, the reason the resulting sets form a partition is that $N_i \cap N(U \setminus \{i\}) =
    \emptyset$ by the definition of $\partial U$. 
    
    For every $Y \subseteq [n]$ if $|N(i)\setminus Y| \ge \delta \Delta$, then $|N_i \setminus Y| \ge
    |N(i) \setminus Y| - |N(i) \setminus N_i| \ge (\delta - 2\eta)\Delta$. It follows that after removing
    $\Err_{\bm{Y}}$ (vertices for which $|N(i)\setminus \bm{Y}| < \delta \Delta$), every vertex $i$ in
    the set $U \setminus \Err_{\bm{Y}}$ in $G_{\bm{X}}$ has at least $(\delta - 2 \eta)\Delta$ neighbours
    in $N_i \setminus \bm{Y}$. As $N_i$ is a partition, it follows that $|N(U) \setminus \Err_{\bm{Y}}
    \setminus N(\Err_{\bm{Y}})| \ge (\delta -2\eta) \Delta |U \setminus \Err_{\bm{Y}}|$.

    Finally, choosing $\eta = \delta / 4$ completes the proof. By \cref{lm:exp-parameters} this particular choice only affects the hidden
    constant in $r = \Omega(n / \Delta)$. 

\section{Proof of \cref{lem:expander-closure}}
\label{sec:expander-proof}
Since $\Cl^X$ and $\Cl^Y$ are independent of each other, we just focus on constructing $\Cl^X$. It suffices to prove the following lemma:
\begin{lemma}
\label{lem:tech-exp-closure}
    Let $G = ([m], [n], E)$ be an $(r, \Delta, \alpha\Delta)$-expander. Let $\mathcal{T}$ be a tree with
    nodes labeled with subsets of $[n]$, where $S_v \subseteq [n]$ denotes the label of $v$ such that 
    \begin{itemize}
        \item For the root of $\mathcal{T}$, the node $r$ we have $S_r = \emptyset$.
        \item If $u$ is a parent of $v$, then $S_u \subseteq S_v$. 
        \item For every $u$ we have $|S_u| \le d \le (\alpha - \beta)^2r \Delta / 4$.
    \end{itemize}
    Then there for every node $u$ to $\mathcal{T}$ there exists a set $T_u \subseteq [m]$ such that 
    \begin{enumerate}[label=(\alph*)]
        \item The graph $G_u \coloneqq G - T_u - S_u - N(T_u)$ is an $(r, \Delta, \beta\Delta)$-expander. \label{item:expansion}
        \item If $u$ is a parent of $v$, then $T_u\subseteq T_v$.
        \label{item:subset}
        \item $|T_u| \le \frac{1}{\alpha - \beta}d/\Delta$. 
        \label{item:size}       
    \end{enumerate}
\end{lemma}
To finish the proof of \cref{lem:expander-closure} given \cref{lem:tech-exp-closure} we just let $\mathcal{T}$ be the tree of the protocol and $S_u$ be $\fix(X_u)$, then take $\Cl^X(u) \coloneqq T_u$.

We now proceed to prove \cref{lem:tech-exp-closure}.
Wlog we may assume that if $u$ is a parent of $v$ we have $|S_v \setminus S_u| \le 1$ (just by replacing a single edge in $\mathcal{T}$ by a chain of edges). 

We construct the sets $T_u$ inductively starting from the root $r$ where $T_r = \emptyset$. Suppose $u$ is a parent of node $v$ and we have constructed $T_u$. If $S_u = S_v$, we just let $T_v \coloneqq T_u$, so assume that $S_v \setminus S_u = \{i\}$. Let $G'_u \coloneqq G_u - i$. Let us find the largest set $B_v \subseteq [m] \setminus T_u$ such that $|B_v| \le r$ and $|N_{G'_u}(B_v)| \le \beta \Delta |B_v|$ and let $T_v \coloneqq T_u \cup B_v$. Then $G_v = G'_u - T_u - N_{G'_u}(T_u)$. It is clear that $T$ satisfies \Cref{item:subset}.

\paragraph{Proof of \Cref{item:size}} We show by induction on the depth $\ell$ of a node $u$ that $|T_u| \le \frac{1}{\alpha - \beta} \ell / \Delta$. The base case is satisfied since for the root $r$ the set $T_r$ is empty. Now let $u$ be a node at depth $\ell$ and $v$ be its child at depth $\ell+1$. We have that $|T_u| \le \frac{1}{\alpha - \beta} \ell / \Delta$, we need to prove that $|T_v| = |T_u \sqcup B_v| \le \frac{1}{\alpha - \beta} (\ell + 1) / \Delta$. 

On the one hand $N_{G'_u}(B_v) = N_G(B_v) \setminus (N_G(T_u) \cup S_v)$. On the other hand, $|N_{G'_u}(B_v)| \le \beta \Delta |B_v|$. By the expansion of $G$ we have $|N_G(B_v)| \ge \alpha \Delta |B_v|$. Hence $|N_G(T_u) \cup S_v| \ge (\alpha - \beta) \Delta |B_v|$. By the assumption on the tree $|S_v| = \ell + 1$, and by induction hypothesis $|T_u| \le \frac{1}{\alpha - \beta} \ell / \Delta$, so $|N_G(T_u))| \le \frac{1}{\alpha - \beta} \ell$.

Combining the two inequalities, we get $\frac{1}{\alpha - \beta}\ell + (\ell + 1) \ge (\alpha - \beta)\Delta |B_v|$.

From that, we get $|B_v| \le 2 \cdot \frac{1}{\left(\alpha - \beta\right)^2\Delta} \cdot (\ell + 1) \le r/2$, where the last inequality follows from the assumptions on $\ell$. Then we get that $|T_v| \le |T_u| + |B_v| \le r$. Now we can use expansion of $G$ to bound $|N_G(T_v)| \ge \alpha \Delta |T_v|$. On the other hand, let $r = w_0, w_1, \dots, w_\ell = u, w_{\ell+1} = v$ be the path in $\mathcal{T}$ from the root to $v$. We then have
\[ N_G(T_v) \subseteq \bigcup_{i = 0}^{\ell} N_{G'_{w_i}}(B_{{w_{i+1}}}) \cup S_v.\]
By the choice of sets $B$ we get $|N_G(T_v)| \le \beta \Delta |T_v| + |S_v|$. Combining the two bounds we get $|T_v| \le \frac{1}{\alpha - \beta} |S_v| / \Delta$, which concludes the proof. 

\paragraph{Proof of \Cref{item:expansion}} Pick the node $v$ at depth $\ell + 1$ such that $G_v$ is not an $(r,\Delta, \beta\Delta)$-expander, and $v$ is the closest to the root among such nodes. In particular, for its parent $u$ the graph $G_u$ is $(r,\Delta,\beta\Delta)$-expander. Then there exists a set $T$ of size at most $r$ such that $N_{G_{v}}(T) < \beta \Delta |T|$. By expansion of $G$ we get $|N_G(T)| \ge \alpha \Delta |T|$. Then, since $N_{G_v}(T) = N_G(T) \setminus (N_G(T_v) \cup S_v)$ we have
\[ \frac{1}{2}(\alpha - \beta)\Delta r \ge \frac{2}{\alpha - \beta}\ell \ge \frac{1}{\alpha - \beta }\ell + (\ell + 1) \ge |N_G(T_v) \cup S_v| \ge (\alpha - \beta)\Delta |T|.  \]
The left-hand side follows from \Cref{item:size} and the right-hand side follows from the analysis above. Then $|T| \le r/2$. Since by the proof of \Cref{item:size} we have that $|B_v| \le r/2$, we get $|T \cup B_v| \le r$, yet $|N_{G'_u}(T \cup B_v)| < \beta \Delta |B_v| + \beta \Delta |T| \le \beta \Delta |B_v \sqcup T|$, contradicting the choice of $B_v$.

\section{Proof of \Cref{lemma:concentration}}\label{sec:concentration_proof}
\LemmaConcentration*

\begin{proof}
    Let $\lambda \in (0, \zeta)$ be some parameter that will be determined later.
    First, observe that for any $1 \le k \le n$ and $a_1, \ldots, a_{k - 1} \in \mathbb{R}$ such
    that $\Pr[\bm{a}_1 = a_1, \ldots, \bm{a}_{k - 1} = a_{k - 1}] > 0$,
    \begin{equation}
        \label{eqn:expt_exp}
        \Exp[\exp(\lambda\bm{a}_k) \mid \bm{a}_1 = a_1, \ldots, \bm{a}_{k - 1} = a_{k - 1}] \le
        \zeta \int\limits_{0}^{\infty} \exp(\lambda x) \cdot \exp(-\zeta x) \cdot \dd{x} =
        \zeta / (\zeta - \lambda).
    \end{equation}

    Next, we prove by induction on $k$ from $n$ to $1$ that
    \begin{equation}
        \label{eqn:induction_hyp}
        \Exp\left[ \exp\left(\lambda\sum_{i = k}^n \bm{a}_i \right) \middle| \bm{a}_1 = a_1, \ldots,
          \bm{a}_{k - 1} = a_{k-1} \right] \le
        \left( \frac{\zeta}{\zeta - \lambda} \right)^{n - k + 1}.
    \end{equation}

    The base case $k = n$ is exactly $\eqref{eqn:expt_exp}$.
    Now assume that \eqref{eqn:induction_hyp} holds for all $k \ge m + 1$. Then
    \begin{align*}
      &\Exp\left[\exp(\lambda \sum_{i = m}^n \bm{a}_i) \middle| \bm{a}_1 = a_1, \ldots, \bm{a}_{m - 1} =
        a_{m - 1} \right]\\
      \le & \sum_{a_m} \Pr[\bm{a}_m = a_m \mid \bm{a}_1 = a_1, \ldots, \bm{a}_{m - 1} = a_{m - 1}] \cdot
            \exp(\lambda a_m) \cdot \\
          & \hspace{2cm}
            \cdot \Exp\left[\exp(\lambda \sum_{i = m + 1}^{n} \bm{a}_i) \middle| \bm{a}_1 = a_1, \ldots,
            \bm{a}_m = a_m\right]\\
      \le & \left( \frac{\lambda}{\zeta - \lambda} \right)^{n - m} \cdot
            \Exp[\exp(\lambda \bm{a}_m) \mid \bm{a}_1 = a_1, \ldots, \bm{a}_{m - 1} = a_{m - 1}]\\
      =& \left( \frac{\lambda}{\zeta - \lambda} \right)^{n - m + 1}.
    \end{align*}

    Finally, by setting $\lambda = \zeta / 2$, we conclude that
    \begin{align*}
        \Pr\left[\sum_{i = 1}^{n}\bm{a}_i \ge \frac{4}{\zeta} n \right]
           & = \Pr\left[ \exp\left(\lambda\sum_{i = 1}^n \bm{a}_i \right) \ge
             \exp\left(\frac{4\lambda}{\zeta} n\right)\right]\\
           & \le \Exp\left[\exp(\lambda\sum_{i = 1}^n \bm{a}_i)\right] \cdot \exp(-2n)\\
           & \le \left( \frac{\zeta}{\zeta - \lambda} \right)^n \cdot \exp(-2n)\\
           & \le \exp(-n).\qedhere
    \end{align*}

\end{proof}

\section{Proof of \cref{fc:deficiency}}
\label{sec:fc-deficiency}

Follows from the computation:
\begin{align*}
  \Dinf(v) - \Dinf(u) &= -|\fix(X_v)| + |\fix(X_u)| + \Hinf(X_v) - \Hinf(X_u) \\
                      &= -n_k(x,y) + (\Hinf(X_v)-\Hinf(X_u^b))+(\Hinf(X_u^b)-\Hinf(X_{v})) \\
                      &\le -n_k+\log (1/q_u^b)+\left(\gamma\cdot n_k(x,y)+\log (1/p_u^{b,\ge i})\right) &
                                                                                                          \text{(from \cref{lemma:density_restoring_partition})}\\
                      &=-(1-\gamma)n_k(x,y)+h_k(x,y).\qedhere
\end{align*}

\end{document}